\documentclass[a4paper]{iopart}

\usepackage{setstack}
\usepackage{iopams}
\usepackage{graphicx}   

\begin{document}

\title{Collapse and revival in inter-band oscillations of a two-band Bose--Hubbard model}

\author{Patrick Pl\"otz$^1$, Javier Madro\~{n}ero$^2$, and Sandro Wimberger$^1$}
\date{\today}
\address{ $^1$ Institut f\"ur Theoretische Physik, Universit\"at Heidelberg, Philosophenweg 19, 69120 Heidelberg, Germany}
\address{$^2$ Physik Department, Technische Universit\"at M\"unchen, James-Franck-Str. 1, 85748 Garching, Germany}

\begin{abstract}
We study the effect of a many-body interaction on inter-band oscillations in a two-band Bose-Hubbard model with external Stark force. Weak and strong inter-band oscillations are observed, where the latter arise from a resonant coupling of the bands. These oscillations collapse and revive due to a weak two-body interaction between the atoms.
Effective models for oscillations in and out of resonance are introduced that provide predictions for the system's behaviour, particularly for the time-scales for the collapse and revival of the resonant inter-band oscillations. 
\end{abstract}



\emph{Introduction. } 
Recent experiments proved the possibility to study the coherent dynamics of interacting many-particle systems~\cite{interactioninduced_interference, newBlochexp,longBO}. Such realisations of many-body systems with ultra-cold gases in optical lattices have a short but impressive history and open immense possibilities for various fields of physics~\cite{BlochZwergerReview, Bloch}. The demonstration of the well-known phenomenon of collapse and revival, the latter being a pure quantum effect, with ultra-cold atoms bears witness of this coherent evolution of a many-body wave function~\cite{newBlochexp,GreinerCR}. Additionally, the high degree of control in such experiments allows a manipulation of many system parameters and makes them particularly interesting for various fields of physics as well as future applications~\cite{Bloch, Application}. Different ways of addressing additional degrees of freedom in such ultra-cold bosonic gases have been suggested~\cite{Bloch}. 

In the present Fast Track Communication, we discuss a two-band model with an additional external force to control the coupling between the two bands. Applying a force to atoms in optical lattices leads to Bloch oscillations and is a realization of a many-body Wannier--Stark system~\cite{morsch}. The coupling of the low-lying energy bands in such systems has been demonstrated in different experiments, e.g. on Landau-Zener tunnelling~\cite{LZ, morsch} and the influence of the many-body interaction on a weak coupling of the bands has also been studied theoretically~\cite{TMW2}. We go beyond a weak coupling of energy bands and consider an isolated two-band system with a strong external force.
A closed two-band system is an idealisation but can be realised approximatively with ultra-cold atoms~\cite{BlochZwergerReview} using different techniques as, e.g., super  lattices~\cite{WeitzMinibands, Witthaut}. Besides the possibility of experimental realisation, a closed two-band model is also interesting as a simple model system. For the latter, we focus on the two lowest energy bands of interacting bosons in a optical lattice $V(x) = V_0\cos (2 k_L x)$, with the wave vector of the optical lattice $k_L = 2\pi/\lambda_L$. Then, all parameters of the model Hamiltonian just depend on this external potential. The parameters can be computed numerically (see, e.g., Appendix A in~\cite{TMW1}) and analytical approximations exist for them for not too small amplitudes $V_0$~\cite{BlochZwergerReview}. Using this setup, we are able to identify regions of strong and weak inter-band coupling. A weak two-body interaction introduces new energy scales in the coherent evolution of the many-body wave function, leading to a collapse of such oscillations, but the form of the interaction gives rise to subsequent revivals. We give analytical expressions for all time-scales in this many-body realisation of a two-band collapse and revival phenomenon, and experimental ramifications for a realisation with ultra-cold bosonic gases are discussed.

\emph{The many-particle model.~} 
We study a two-band Bose-Hubbard model with an additional external Stark force for a strong coupling of the two bands, introduced in \cite{TMW1, Andrea}

\begin{equation}\label{eq:fullHamiltonian}
\eqalign{
\mathcal H = \sum_{l=1}^L \Big[\epsilon_l^- n_l^a - \frac{t_a}{2}(a_{l+1}^{\dag}a_l^{} + { \rm h.c.}) + \frac{gW_a}{2} n_{l}^a (n_l^a-1)  \cr 
 + \epsilon_l^+ n_l^b + \frac{t_b}{2}(b_{l+1}^{\dag}b_l^{} + { \rm h.c.}) + \frac{gW_b}{2} n_{l}^b (n_l^b-1) + FC_0(b_l^{\dag}a_l +  {\rm h.c.}) \cr
+ 2gW_x n_{l}^a n_l^b + \frac{gW_x}{2}(b_l^{\dag}b_l^{\dag}a_l a_l + {\rm h.c.} ) \Big]. }
\end{equation}
Here, $a_l$ ($a_l^{\dag}$) annihilates (creates) a particle at site $l$ of totally $L$ sites in the lower band and $b_l$ ($b_l^{\dag}$) in the upper band. The corresponding number operators are $n_l^a = a_l^{\dag}a_l^{}$, $n_l^b = b_l^{\dag}b_l^{}$. The bands are separated by a bandgap $\Delta$ and have onsite-energies $\epsilon_l^{\pm} = \pm\Delta/2+lF$, respectively. We include hopping between neighbouring sites in band $a, b$ with a hopping strength $t_a,t_b>0$, and a repulsive interaction between particles occupying the same site in band $a\;(b)$ with a strength $W_a\;(W_b)$. The two bands are coupled via $C_0 F$, with the external Stark force $F$ and a coupling constant $C_0$ depending on the depth of the lattice $V_0$~\cite{TMW1, Andrea}, and also via the inter-particle interaction with a strength $W_x$. All parameters are measured in recoil energies $E_{\rm rec}\equiv \hbar^2k_L^2/(2m)$ and we set $\hbar = 1$ troughout. Focussing on a realisation with a single optical lattice rather than a superlattice, the relation between the parameters is generally: $\Delta\gg t_a, t_b$, as well as $t_a,t_b\approx W_i$ and $C_0=\mathcal O(10^{-1})$. We take the external force $F$ as a free parameter.
We assume that the interaction strength can be tuned (experimentally, e.g, by the use of Feshbach resonances~\cite{BlochZwergerReview}) and include a corresponding scaling factor $g$ to all interaction terms. 
To study the occupation of the upper band, we prepare the system in an initial state $|\psi(0)\rangle$, with a uniform distribution of particles in the lower band only and evolve it in time by the many-body Schr\"odinger equation. The quantity we study is the (normalised) number of particles in the upper band
\begin{equation}
	\mathcal N_b (t) \equiv \frac{1}{N} \langle\psi (t)| \sum_l n_l^b |\psi(t)\rangle,
\end{equation}
where $N = \sum_l (n_l^a+n_l^b)$ is the total number of particles. We will refer to $\mathcal N_b(t)$ as \emph{occupation of the upper band}. 

Let us discuss the non-interacting single-particle case $\mathcal H_0 \equiv \mathcal H (g=0)$ first. We apply the following trans\-for\-mat\-ion involving Bessel functions of the first kind $J_n(x)$ to our operators, which is known to remove the hopping terms in the single-band case~\cite{Fukuyama} 
	\begin{equation}
		\alpha_n  = \sum_{l\in\mathrm Z} J_{l-n}(x_a) a_l \qquad \beta_n  = \sum_{l\in\mathrm Z} J_{l-n}(x_b) b_l,
	\end{equation}
with $x_{i} \equiv t_{i}/F$, $i = a,b$. Using relations for Bessel functions, we arrive at 
	\begin{equation}\label{eq:transformedH}
		\mathcal H_0 = \sum_{l\in\mathrm Z}\Big[ \epsilon_l^{-} \alpha_l^{\dagger}\!\alpha_l +\epsilon_l^{+}\beta_l^{\dagger}\!\beta_l +C_0F \sum_{n} J_{l-n}(\Delta x) ( \alpha_l^{\dagger}\!\beta_{l+n}  + \rm{h.c.} ) \Big],
	\end{equation}
where $\Delta x = x_a + x_b$ and $\epsilon_l^{\pm} = \pm\Delta/2+lF$ as above. We obtain coupling between any two sites of the two different bands, weighted by Bessel functions. In \eref{eq:fullHamiltonian} the coupling between different and possibly remote sites originates from an on-site coupling and subsequent hoppings. It can thus be considered a higher order process in the original basis. But in the transformed Hamiltonian~\eref{eq:transformedH}, this coupling is now direct with a strength modified by the factor $J_{l-n}(\Delta x)$.

The Hilbert space in the many-particle problem is spanned by Fock states with fixed particle number $|n_1^{a}, \ldots, n_{L}^a; n_1^b, \ldots, n_L^b\rangle$ with single-particle basis $a_l, b_l$. The total dimension of the Fock space for a given number of atoms $N$ and lattice sites $L$ per band is given by $\dim \mathcal H = (N+2L-1)!/[N!(2L-1)!]$. For numerical simulations, we change to the interaction-picture with respect to the external force~\cite{FB} which removes the tilt $\sum_l ln_l^{a,b}F$ and replaces $a_{l+1}^{\dagger}a_l^{}\rightarrow \rme^{\rmi Ft}a_{l+1}^{\dagger}a_l^{}$ (and likewise for $b_{l+1}^{\dagger}b_l^{}$). The Hamiltonian is then time-dependent with a periodicity of $T_B\equiv 2\pi/F$ and decomposes into a direct sum of operators for specific quasi-momenta $\kappa$~\cite{FB}. As a consequence, the size of the Hilbert space is reduced by a factor of the order of $L$~\cite{TMW1, Andrea,FB}. Since the different subspaces are physically equivalent, we restrain our discussion to the $\kappa = 0$-subspace of the Hilbert space~\cite{FB}.\\
For the time-evolution of a given initial state, we use either a direct numerical integration (with an adaptive step-size Runge-Kutta algorithm~\cite{NR}) or an eigenbasis expansion after diagonalising the problem:
\begin{equation}\label{eq:eigenbasis}
	|\psi(mT_B)\rangle = \sum_n c_n \exp(-\rmi \varepsilon_n mT_B)|\varepsilon_n\rangle. 
\end{equation}
Here we use the eigenstates $U_F(T_B)|\varepsilon_n\rangle = \exp(-\rmi \varepsilon_nT_B)|\varepsilon_n\rangle$ of the Floquet-Bloch operator ($\mathcal T$ denotes time-ordering)
\begin{equation}\label{eq:FBoperator}
	U_F(T_B) = \mathcal T {\rm exp}\Big[-{\rm i} \int_0^{T_B}\mathcal H(t) {\rm d}t\Big],
\end{equation}
since the Hamiltonian is explicitly time-dependent, with a periodicity $T_B$. In addition, this gives us the full spectrum and enables access to relevant energy scales of the problem, as well as an identification of the most important states participating in 
the time-evolution.

\begin{figure}[th!]
	\begin{center}
	\includegraphics[width = 0.95\linewidth]{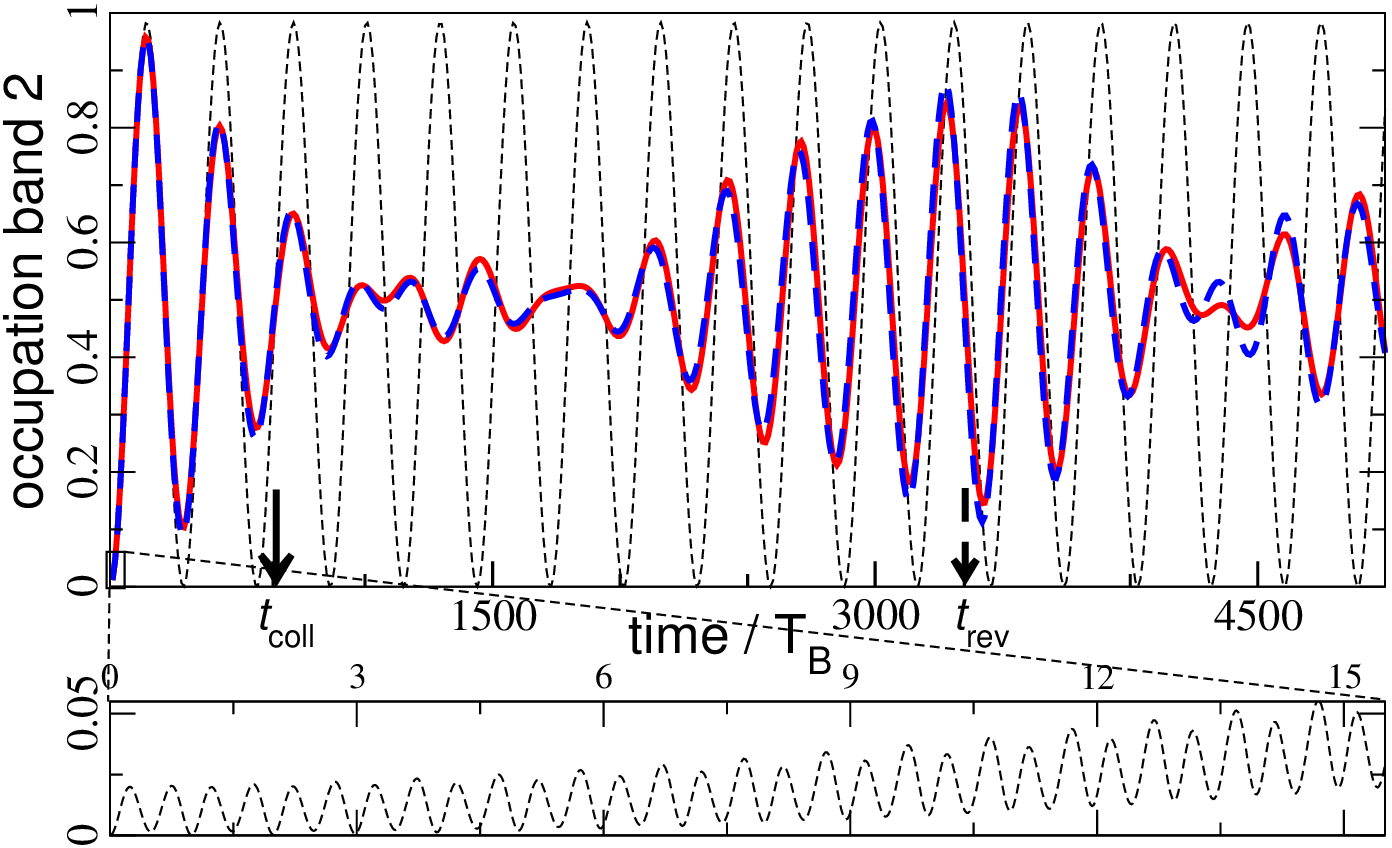}
	\caption{\label{fig1} (colour online) Occupation of the upper band as a function of time for a resonance of order $r=2$. Shown are the cases: vanishing two-body interaction ($g=0$, thin dashed line), weak two-body interaction ($g=0.1$) with all interaction terms (thick line) and only one interaction term $2W_x\sum_l n_l^a n_l^b$ (thick dashed line) included. 
	The collapse and revival times, $t_{\rm coll}$ and $t_{\rm rev}$, are indicated by arrows. \emph{Lower panel:} Magnification of the initial oscillation showing small non-resonant oscillations of period $T_{\tilde\Delta}$ on top of the resonant oscillations with a much longer period $T_{\rm res}$. Parameters correspond to $V_0 = 4$: $\Delta = 4.39, C_0 = -0.15, t_a = 0.062, t_b = 0.62, W_a = 0.030, W_b = 0.018, W_x = 0.012; N = L = 5$ and $F = 2.2207$.}
	\end{center}
\end{figure}

\emph{Results.~}
For the specific system under consideration, i.e., bosons in optical lattices, both hopping coefficients $t_a, t_b$ are smaller than unity and (since we are interested in strong inter-band coupling) we take the external Stark force $F$ to be 
much larger than the hopping coefficients: $\Delta x \equiv (t_a+t_b)/F\ll 1$. 
The non-interacting Hamiltonian \eref{eq:transformedH} now allows for simple solutions for two regimes: values of the external force $F$ not leading to a degeneracy between energy levels of different bands (\emph{off-resonant} regime) and values of the force $F$ leading to such a degeneracy (\emph{resonant} regime). For the off-resonant regime we make use of $\Delta x\ll 1$ and neglect all Bessel functions in \eref{eq:transformedH} except for $J_0(\Delta x) \approx 1$. The Hamiltonian then decomposes~\cite{Nakamura} into a sum of independent two-level systems and the occupation of the upper band when initially zero follows a simple Rabi formula $\mathcal N_b(t) = [1+\tilde\Delta^2/(4C_0^2F^2)]^{-1}\sin^2 (\tilde\Delta t/2)$, where $\tilde\Delta \equiv\sqrt{\Delta^2+4C_0^2F^2}$. This corresponds to Rabi oscillations between the bands with an amplitude much smaller than unity and a period $T_{\Delta} = 2\pi/\tilde \Delta $ of the order of the \emph{Bloch period} $T_B = 2\pi/F$. An example is shown in the lower panel of \fref{fig1} where the off-resonant contribution to the oscillations is shown. 

\noindent Although the coupling from a site $l$ to sites in the other band with different index $l'$ is usually small (cf. discussion after \eref{eq:transformedH}), it is important when the two levels become degenerate in energy, i.e., for resonant values of the force $F$. This happens when the energy gap between both bands is close to an integer multiple of the external force $\tilde\Delta \approx rF$ 
and we refer to this regime as \emph{resonant of order $r$}.
In resonance, the coupling of the degenerate levels is most important and the Hamiltonian of \eref{eq:transformedH} can similarly be reduced to a sum of independent two-level systems
	\begin{equation}\label{eq:resonantH}
		\mathcal H_0 = \sum_{l\in\mathrm Z} \Big[ \epsilon_l^{-} \alpha_l^{\dagger}\!\alpha_l +\epsilon_l^{+}\beta_l^{\dagger}\!\beta_l +C_0F J_{l-r}(\Delta x) ( \alpha_l^{\dagger}\!\beta_{l+r}  + \rm{h.c.} ) \Big],
	\end{equation}
and diagonalised by $\mu_l^{(r)} = \frac{1}{\sqrt{2}}(\beta_l+\alpha_{l+r}^{}) $ and $\nu_l^{(r)} = \frac{1}{\sqrt{2}}(\beta_l-\alpha_{l+r}^{})$. The resonant oscillations (of order $r$) between the two bands have an amplitude of almost unity and a period given by $T_{\rm res} = \pi/|C_0FJ_r(\Delta x)| \gg T_B. $
An example of these resonant oscillations for $r=2$ is shown as the thin dashed line in \fref{fig1}. The period predicted by the reduction to independent two-level systems is $T_{\rm res}^{r=2} = 285\; T_B$ for the parameters given there, in excellent agreement with \fref{fig1} where the actual period is $T_{\rm res} \approx 288\; T_B$. We found equally good agreement in numerical simulations for different lattice depths and other orders $r$ of resonance not explicitly reported here. 

\begin{figure}[ht!]
	\begin{center}
	\includegraphics[width = 0.95\linewidth]{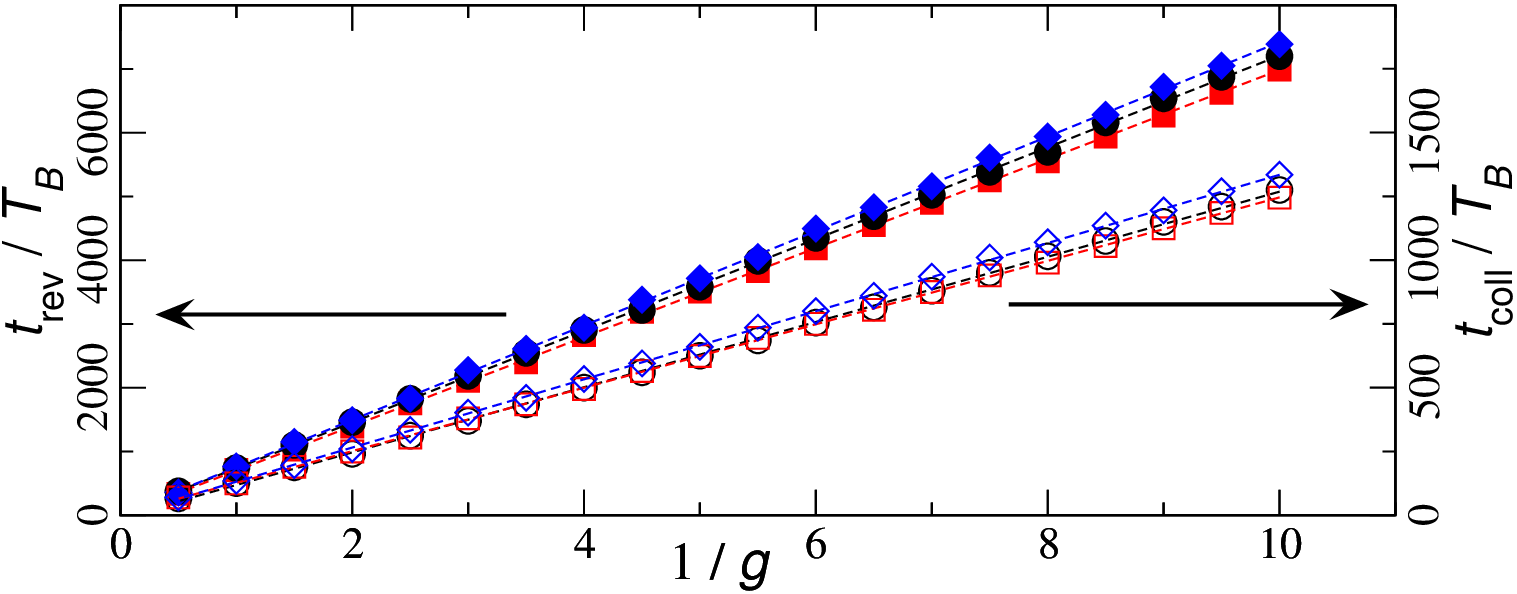}
	\caption{\label{fig2} (colour online) Collapse times (open symbols) and revival times (filled symbols) versus 
	$1/g$ for $N,L=5,5$ and $V_0=4$: \opensquare, $V_0 = 5$: \opencircle, $V_0 = 6$: \opendiamond. Order of the resonance: $r=1$. The collapse time is defined via $\mathcal N_b(t_{\rm coll}) = (1+\rme^{-1})/2$ and the revival time is chosen as the next maximum in the oscillations after initial decay $t> t_{\rm coll}$, as indicated in \fref{fig1} (upper panel). 
	}
	\end{center}
\end{figure}

Let us now study the effect of the many-body interaction in the original Hamiltonian~\eref{eq:fullHamiltonian}. \Fref{fig1} also shows the occupation of the upper band as a function of time for the initial state $|\psi(0)\rangle = |1,\ldots,1; 0,\ldots,0\rangle$ in a weakly interacting system with $g=0.1$. We observe a decay of the resonant oscillations followed by a major revival. At later times several minor revivals occur (not shown in \fref{fig1}). This effect is stable against variations of the system parameters (as number of particles and lattice sites, even for fillings $N/L$ of order but not always close to one), with the time scales of decay and revival depending on the interaction strength $g$ (see the following section and \fref{fig2}). We found the same phenomenon with different initial states in numerical simulations as long as the particles are mainly delocalised along the lattice and occupy only the lower band, i.e., excluding Fock states with all particles on one lattice site for instance.

The oscillatory behaviour depicted in \fref{fig1} strongly reminds of the collapse and revival effect known from quantum optics~\cite{Meystre}. A specific feature in these systems is a linear dependence of the collapse and revival time on the inverse coupling strength, i.e. $t_{\rm rev}\propto 1/g$ and $t_{\rm coll}\propto 1/g$ where $g$ usually denotes the coupling strength between the light field and the two-level atom in the quantum optical systems~\cite{Meystre}. We verified numerically for our system that the observed decay and revival times obey a similar dependence with the interaction parameter $g$ as coupling strength in our model. We define $t_{\rm coll}$ as the time when the difference between the maximal and average amplitude of the inter-band oscillations has fallen to $1/\rme$, i.e., $\mathcal N_b(t_{\rm coll})\equiv 1/2+1/(2\rme)$. The revival time is taken as the maximum of the revived oscillations. \Fref{fig2} clearly demonstrates the linear dependence of these times on the inverse interaction strength.

To find an effective description, we try to understand the most relevant interaction processes. Clearly, the repulsive two-body interaction disfavours double occupancy of sites since this will always cost an interaction energy $W_{a(b)}$ for two particles occupying the same site in the lower (upper) band. 
Starting from an initial state with population of the ground band only, the strong Stark force leads to an occupation of the upper band. Doubly occupied sites are also suppressed there, but two particles may sit at the same site in either band, i.e. ``on top of each other". Indeed, the most important interaction term in the time-evolution is $2W_x\sum_l n_l^an_l^b$, since it already gives a non-zero contribution when there is only one particle per site in each of the two bands. In fact, comparing the time evolution of the initial state with all interaction terms and only the one mentioned shows almost no difference (cf. \fref{fig1}). We focus on fillings close to unity $\bar n = N/L \approx1$ and study the time-evolution of states of the form $|\psi_0\rangle = |1,1,\ldots,1; 0,\ldots,0\rangle$, which is not an eigenstate of the system in resonance. 
Note that this is the most important contribution to the superfluid ground state in an expansion in our configuration state Fock basis. For large enough systems the superfluid ground state of the untilted system (and the single-occupancy state $|\psi_0\rangle$ likewise) becomes indistinguishable from a coherent state (eq.~(66) in~\cite{BlochZwergerReview}) that factorises into a product of local coherent states at each site $l$:
\begin{equation}\label{eq:cohstate}
	\prod_l \Big( \rme^{\sqrt{\bar n} a_l^{{\dagger}}} |{\rm vac}\rangle_l\Big) = 
	\prod_l |\varphi; 0\rangle_l 
	\equiv |\boldsymbol \varphi; 0\rangle. 
\end{equation}
We denote this coherent state with phase $\varphi = \sqrt{\bar n}$ by $|\boldsymbol \varphi; 0\rangle$. 
We are now going to re-write this state in the resonant basis and determine the effect of the interaction term $2W_x\sum_ln_l^an_l^b$ when acting on this state. We start by inserting the transformation $a_l^{\dagger}= (1/\sqrt{2})\sum_nJ_{l-n}(x_a)(\mu_n^{\dagger}- \nu_n^{\dagger})$ into \eref{eq:cohstate} to obtain
\begin{eqnarray}\label{eq:rescoherent}
	|\varphi; 0\rangle & =  \prod_l {\rm exp}\bigg[ \sqrt{\frac{N}{2L}}\sum_nJ_{l-n}(x_a)(\mu_n^{\dagger}- \nu_n^{\dagger}) \bigg] |{\rm vac}\rangle \nonumber\\
	& = \prod_n \rme^{\sqrt{\bar n/2}\;\mu_n^{\dagger}} \rme^{-\sqrt{\bar n/2}\;\nu_n^{\dagger}} |{\rm vac}\rangle. 
\end{eqnarray}
where we used $\sum_{m\in\mathrm Z }J_m(x)z^m = {\rm exp}[x(z-1/z)/2]$ for $z=1$~\cite{AS} and $[\mu_n^{\dagger},\nu_n^{\dagger}]=0$ which follows from the properties of the operators $a^{(\dagger)}, b^{(\dagger)}$. Let us denote Fock states with single-particle basis $\mu_l, \nu_l$ by round parentheses $|n_1^{\mu}, \ldots, n_{L}^{\mu}; n_1^{\nu}, \ldots, n_L^{\nu} )$. From \eref{eq:rescoherent} we see, that the coherent state of the lower band in resonance is a product of local coherent states for both bands in the resonance basis $\prod_l|\tilde\varphi; -\tilde\varphi)_l\equiv|\boldsymbol{\tilde\varphi}; \boldsymbol{-\tilde\varphi})$ with $\tilde\varphi\equiv \varphi/\sqrt{2}$. The time-evolution of this state (for the non-interacting system in resonance) is simple since it is diagonal in the eigenbasis of the Hamiltonian in resonance \eref{eq:resonantH}.
We continue to study the phase evolution created by the most important term $2W_x\sum_ln_l^an_l^b$ perturbatively by expressing the operators $a_l^{(\dagger)},b_l^{(\dagger)}$ in the $\mu,\nu$-basis:
\begin{eqnarray}\fl
	\exp\Big[\rmi 2gW_xt\sum_ln_l^an_l^b\Big] |\boldsymbol{\tilde\varphi}; -\boldsymbol{\tilde\varphi}) 
	 = \exp\bigg[\frac{\rmi gW_x t }{2}\sum_{l}\sum_{l_1,\ldots,l_4} J_{l-l_1}(x_a)J_{l-l_2}(x_a)J_{l-l_3}(x_b) \times \nonumber\\
	J_{l-l_4}(x_b) (\mu_{l_1}^{\dagger} - \nu_{l_1}^{\dagger})(\mu_{l_2} - \nu_{l_2})(\mu_{l_3}^{\dagger}+ \nu_{l_3}^{\dagger})(\mu_{l_4} + \nu_{l_4})\bigg] |\boldsymbol{\tilde\varphi}; -\boldsymbol{\tilde\varphi}).
\end{eqnarray}
Since $|\boldsymbol{\tilde\varphi}; -\boldsymbol{\tilde\varphi})$ is a product of local coherent states we can ignore the sum over $l$ and discuss the expected behaviour. Firstly, both $x_{a,b}$ are much smaller than unity and the main contribution in the sums over $l_1,\ldots, l_4$ will come from the zeroth-order Bessel functions $J_0$. Secondly, the product of operators gives $16$ different combinations of the field operators. But due to the prefactors, the combinations with equal indices are the most important. They simply give an integer number when applied to the product of local coherent states they are acting on. Taking these two points together, the time evolution of this state should show an approximate revival at
\begin{equation}\label{eq:revivaltime}
	t_{\rm rev} \approx \frac{4\pi}{gW_x J_0^2(x_a) J_0^2(x_b)}.
\end{equation}
This result is valid for large systems and cannot account for the effect of non-universal properties like a limited number of particles and lattice sites, but we expect it to yield the right order of magnitude for finite systems, and in particular the correct scaling with the parameters of external potentials (c.f. \fref{fig4}, inset, below).

\begin{figure}[ht!]
	\begin{center}
	\includegraphics[width = 0.95\linewidth]{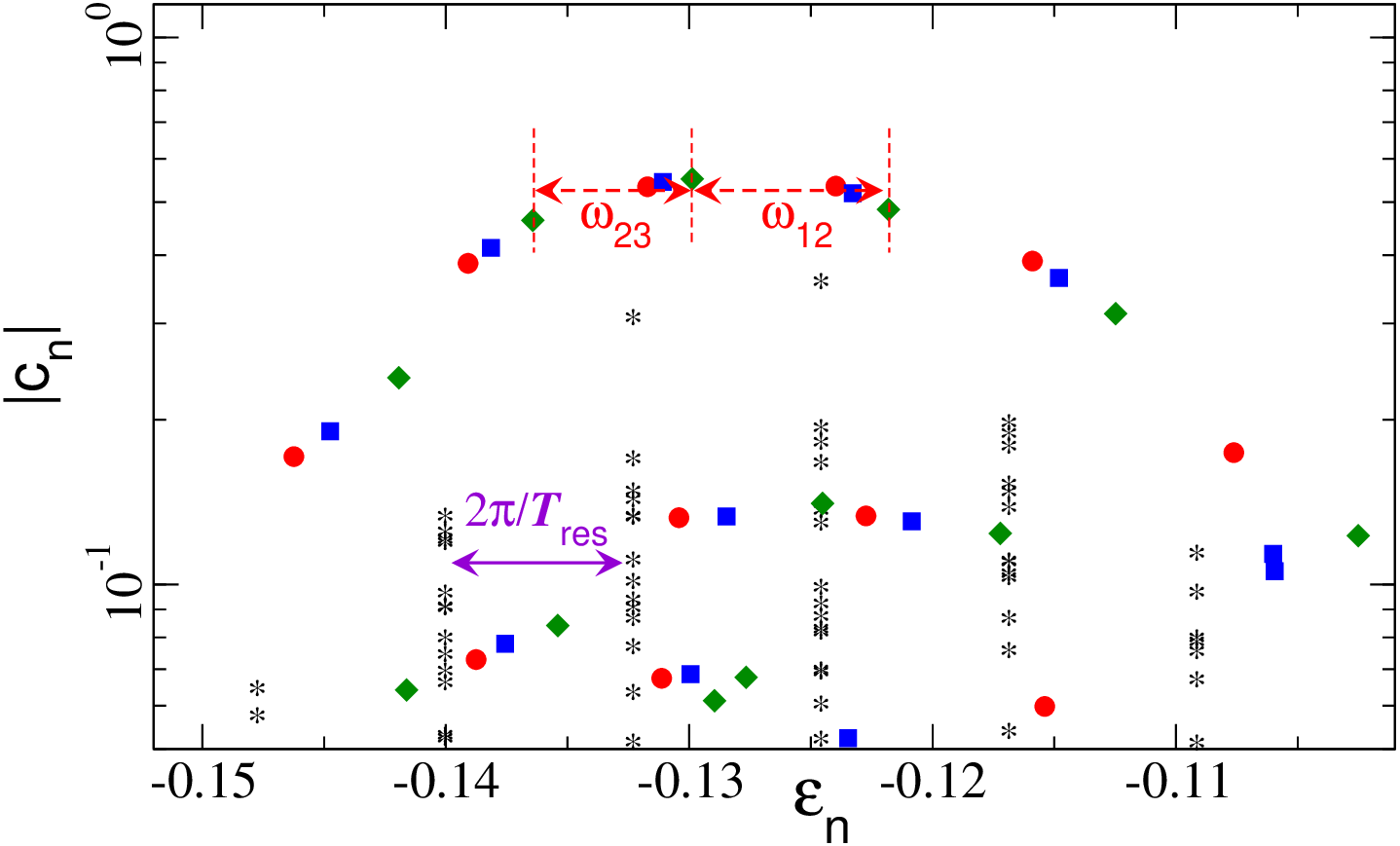}
	\caption{\label{fig3} (colour online) 
	$|c_n|$ versus the corresponding quasi-energies $\varepsilon_n$ in an expansion in the eigenbasis of the Floquet--Bloch operator \eref{eq:FBoperator}. Shown are different interaction strengths: $g=0.0$ (*), $g = 0.05$ (\fullcircle$\!$), $g = 0.1$ (\fullsquare), and $g = 0.2$ (\fulldiamond); other parameters as in \fref{fig1}. We observe that different quasi-energies are shifted at different rates when increasing the interaction strength.}
	\end{center}
\end{figure}

\begin{figure}[ht!]
	\begin{center}
	\includegraphics[width = 0.95\linewidth]{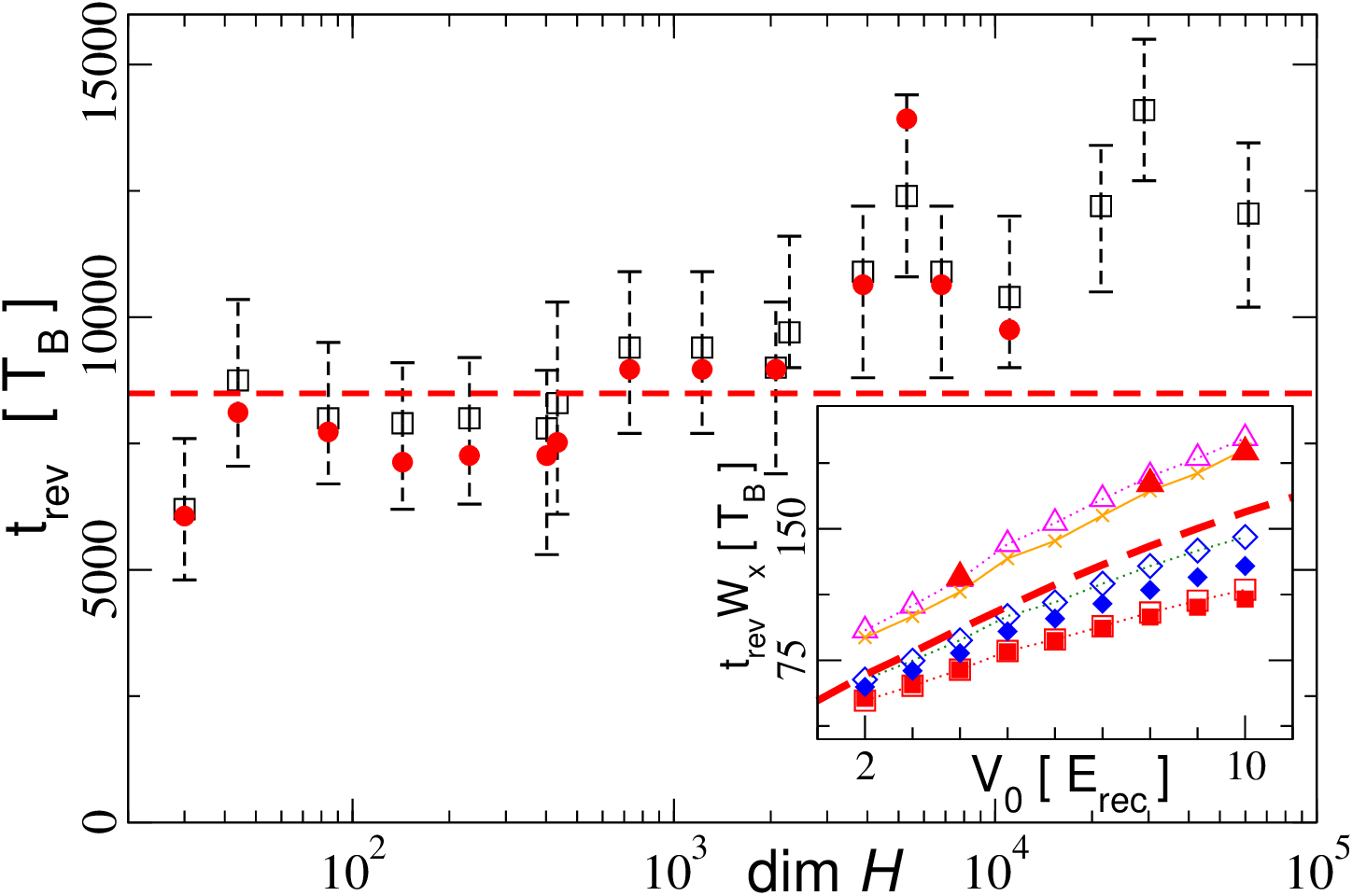}
	\caption{\label{fig4} (colour online) Comparison between estimated revival times according to \eref{eq:revivaltime} (\broken) and to \eref{eq:revivaltime2} (\fullcircle$\!\!$) with numerical simulations (\opensquare) for $V_0=8, r=1$, and different system sizes, parametrised by the Hilbert space dimension of the $\kappa=0$ subspaces. The error bars indicate the width of the revival at half maximum.
	\emph{Inset:} Scaling of the numerically measured revival time with the lattice depth $V_0$ for $r=1$. We multiply $t_{\rm rev}$ by $W_x$ since $W_x$ also depends on $V_0$ to show the remaining non-trivial scaling behaviour. Shown are $N,L=4,4$ ($\dim \mathcal H = 86$, \opensquare), $N,L=5,5$ ($\dim \mathcal H = 402$, \opendiamond), 
	$N,L=6,7$ ($\dim \mathcal H = 3876$, \opentriangle) $N,L=7,7$ ($\dim \mathcal H = 11076$, $\times$). Filled symbols are $t_{\rm rev}$ expected from \eref{eq:revivaltime2} for the same system, and the thick dashed line again shows our universal result \eref{eq:revivaltime}.}
	\end{center}
\end{figure}

Additional finite size corrections to \eref{eq:revivaltime} can be understood by using the decomposition into the eigenbasis~\eref{eq:eigenbasis}. The evolution of the occupation of the upper band certainly depends on the initial state, which we can take into account by studying the weights $c_n$ of the initial state expanded in the eigenbasis. The result of a numerical diagonalisation for a system in resonance is depicted in \fref{fig3}, where the absolute values of the expansion coefficients with their corresponding quasi-energies are shown. 
For vanishing two-body interaction strength $g=0$, the quasi-energies from the states with different occupation numbers are degenerate as expected. The energy difference between neighbouring lines of constant quasi-energies corresponds to the time-scale of the resonant inter-band oscillations and follows from the diagonalisation of the resonant non-interacting system as $\Omega_{\rm res} = 2\pi/T_{\rm res} = 2 | C_0FJ_r(\Delta x)|$. In the non-interacting system ($g=0$), two coefficients are dominating and the difference of the quasi-energies yields a single time-scale $T_{\rm res}$. When the interaction is turned on, the weight of states with many contributions from double- or higher occupancies of sites decrease significantly (since they are energetically disfavoured) and their quasi-energies are slightly shifted. But, surprisingly, only a limited number of additional coefficients $c_n$ contributes significantly in the eigenbasis expansion even for $g\neq0$. The observed collapse and revival signal is now determined by a few expansion coefficients that are much larger than the others. If we focus on the three largest coefficients, denoted as $c_1, c_2,c_3$ and sorted by their quasi-energies $\varepsilon_1, \varepsilon_2, \varepsilon_3$, we find that the latter are shifted by the interaction by different amounts. The differences between neighbouring quasi-energies, $\omega_{12}=|\varepsilon_2-\varepsilon_1|$ and $\omega_{23}=|\varepsilon_3-\varepsilon_2|$ (shown for the example $g=0.2$ in \fref{fig3}), lead to a beating between two oscillations with periods $T_{12},T_{23}\approx T_{\rm res}$ and the revival time will thus be given by
\begin{equation}\label{eq:revivaltime2}
t_{\rm rev} \approx \frac{2\pi}{\omega_{23}-\omega_{12}} = \frac{T_{12}T_{23}}{T_{23}-T_{12}}.
\end{equation}
This estimate requires a numerical diagonalisation but gives a clear physical interpretation to the revival time observed in a specific realisation with $N$ atoms on $L$ lattice sites.
Thus, taking \eref{eq:revivaltime} and \eref{eq:revivaltime2} together, we have an understanding of the general physical mechanism triggering the collapse and revival of the resonant two-band oscillations.  

We compare our prediction for the revival time, \eref{eq:revivaltime} and \eref{eq:revivaltime2}, to actual numerical simulations in \fref{fig4}.
We find that \eref{eq:revivaltime} gives the right order of magnitude for the revival in a specific realisation, and, in particular, it shows the correct scaling behaviour with the depth of the optical lattice (shown in the inset of \fref{fig4}). 
Additionally, \fref{fig4} shows that the corrections for specific sizes of the system are well accounted for by \eref{eq:revivaltime2}, which only slightly underestimate the revival times by a few percent. This deviation could be corrected by including more than three participating states, extending in this manner the arguments which lead to \eref{eq:revivaltime2}.\\
In an experimental realisation, the shorter the collapse time $t_{\rm coll}$ the easier it could be observed, and for an estimate we make use of the fact~\cite{Meystre, Buchleitner_wavepakets} that the collapse time is proportional to the revival time 
\begin{equation}
	t_{\rm coll} = \frac{1}{\pi (\Delta n)^2} \; t_{\rm rev},
\end{equation}
where $\Delta n$ denotes the width of the distribution of coefficients $c_n$ necessary to expand the initial state in the eigenbasis. 
For the specific example of $V_0 = 5, g=0.1, r = 1$ and $N=5=L$ we find $\Delta n \approx \mathcal O (1)$ for the width of the distribution (i.e. the $g\neq 0$ couplings include just one or very few additional states as in the derivation of \eref{eq:revivaltime2} above), such that we estimate $t_{\rm rev}/t_{\rm coll} \approx 3$ compared to $t_{\rm rev}/t_{\rm coll} \approx 5.7\pm 0.1$ found numerically. 

In this work, we focused on realisations with ultra-cold atoms in a tilted periodic potential, and the observed effect can be manipulated by engineering the potential~\cite{interactioninduced_interference, WeitzMinibands} or the two-body interaction~\cite{BlochZwergerReview}. Specifically, the revival time \eref{eq:revivaltime} depends sensitively on parameters as the hopping strength and the external force close to the zeros of the Bessel function in \eref{eq:revivaltime}. This is analogous to already realised manipulations by time-dependent forces as predicted by~\cite{eckardt} and realized in~\cite{arimondo}. The observation of Bloch oscillations over thousands of periods and a fine control on the two-body interaction have already been demonstrated experimentally~\cite{longBO}. Therefore, the collapse and revival of resonant inter-band oscillations predicted here should be accessible in such state-of-the-art experiments. 
We finally remark that the collapse and revival phenomena discussed in this section have their origin (cf. \fref{fig1}) in the degradation (due to interactions) of single particle \emph{inter-band} oscillations. So, even if there are analogies to the collapses and revivals observed in other experiments~\cite{newBlochexp, GreinerCR, BEC1}, the collapse-revival oscillations reported there arise from the interaction within a single-band. Therefore, in contrast to our results, those oscillations would not at all occur when the lower band interaction is suppressed, equivalent to $W_a = 0$ in the model here discussed.

\emph{Summary.~} 
We studied the coupling between two energy bands in a two-band Bose--Hubbard model with an additional tilting force. The force can lead to a strong coupling of the bands and we found strong resonances in the inter-band oscillations in this lattice model. Furthermore, the two-particle interaction leads to a collapse and revival of the resonant inter-band oscillations. Here, we made predictions for the relevant time scales which were verified numerically. \\
A closed two-band system is an idealisation, but it can approximately be realised in various parameter regimes with ultra-cold atoms~\cite{BlochZwergerReview} using different techniques as e.g. super lattices~\cite{WeitzMinibands}. In addition, the use of Feshbach resonances allows a complete control and fine tuning of the interaction strength~\cite{longBO} needed to test our predictions. Our work is a first important step in going beyond ground-band physics and accessing more degrees of freedom in bosonic ultra-cold gases. 

\ack
This work was supported within the framework of the Excellence Initiative by the German Research Foundation (DFG) through the Heidelberg Graduate School of Fundamental Physics (grant number GSC 129/1), the Frontier Innovation Fonds, and the Global Networks Mobility Fund. P.P. is grateful to Andrea Tomadin and acknowledges support from the Klaus Tschira Foundation.

\section*{References}

\end{document}